\documentclass[aps,pra,twocolumn,showpacs,prabib]{revtex4}
\usepackage{graphicx}
\usepackage{amsmath}
\usepackage{amssymb,wasysym}
\usepackage{amsfonts}
\usepackage{bm,verbatim}

\usepackage[ansinew]{inputenc}
\usepackage{graphicx}
\usepackage{color}
\usepackage[colorlinks]{hyperref}

\newcommand{\bea}{\begin{eqnarray}}
\newcommand{\eea}{\end{eqnarray}}
\newcommand{\be}{\begin{equation}}
\newcommand{\ee}{\end{equation}}

\begin{document}
\title{Exact solutions for the dispersion relation of Bogoliubov modes localized near a topological defect - a hard wall - in Bose-Einstein condensate}

\author{Peter V. Pikhitsa}\email{peter@snu.ac.kr}
\affiliation{School of Mechanical and Aerospace Engineering,
Seoul National University, Seoul 151-744, South Korea}

\begin{abstract}
We consider a Bose-Einstein condensate of bosons with repulsion, described by the Gross-Pitaevskii equation and restricted by an impenetrable "hard wall" (either rigid or flexible) which is intended to suppress the "snake instability" inherent for dark solitons. We solve analytically the Bogoliubov - de Gennes equations to find the spectra of gapless Bogoliubov excitations localized near the "domain wall" and therefore split from the bulk excitation spectrum of the Bose-Einstein condensate. The "domain wall" may model either the surface of liquid helium or of a strongly trapped Bose-Einstein condensate. The dispersion relations for the surface excitations are found for all wavenumbers $k$ along the surface up to the "free-particle" behavior $k \rightarrow \infty$, the latter  was shown to be bound to the "hard wall" with some "universal" energy $\Delta$.

\end{abstract}

\pacs{03.75.Lm, 03.75.Kk, 03.65.Ge}

\date{\today}

\maketitle

\section{Introduction}
Initially the Gross-Pitaevskii equation (GPE) was intended to describe structures and excitations in superfluid helium \cite{Ginzburg}, \cite{Pitaevskii1}. Being a non-linear Schr\"{o}dinger equation it possesses a broad spectrum of applications for various non-linear processes in condensed matter such as bright and dark solitons in Bose-Einstein condensate (BEC) and non-linear optics \cite{Pitaevskii1}, as well as the waves of finite amplitude on the surface of liquid \cite{Zakharov}. BEC disturbances known as Bogoliubov excitations \cite{Quantum dark soliton}, \cite{A direct perturbation theory for dark solitons} are described by the eigenmodes of the matrix Bogoliubov-de Gennes  equation (BdGE) that follows from GPE and has become an archetype in many fields ranging from superconductivity \cite{de Gennes} to the gravitational black hole analogy in BEC \cite{Holes1}, \cite{Holes2}. The ubiquitous nature of GPE and BdGE demands rigorous analytical solutions though not many of them have been obtained.

Domain wall solutions of GPE such as 2D dark solitons are known to be unstable except the case of a solid wall. Maybe this explains the fact that the BdG equation has not been paid much attention to find the localized solutions near such walls. Yet even the case of the solid wall deserves investigation as far as it is connected with the generic topic of edge excitations in topological phases. The situation may bear some resemblance to two-band models with Majorana bound states that arise as solutions to three dimensional BdG theories. The gapless modes that propagate along a physical boundary, while they are exponentially localized away from the physical boundary are gapless boundary modes or edge states.

 The surface excitations in restricted BECs (superfluid helium 4 with BEC confined in pores \cite{Shams}, self-bound BEC at the surface of superfluid helium \cite{Griffin}, as well as at the surface of  BEC trapped in an external potential  \cite{Anglin}, or BEC near a solid wall \cite{Kuznetsov}) are of fundamental interest.  One way to obtain surface excitations of BEC was considered in the work by Anglin {\it et al.} \cite{Anglin} which treats the surface excitation of stable BEC (half analytically, half numerically) in the presence of an external linear trapping potential.

 Unlike the "soft" trapping potential \cite{Anglin}, in order to make the problem analytically tractable one may consider the extreme boundary condition of the solid wall  for the surface of trapped BEC which stability was proven in \cite{Kuznetsov} by showing that the imposition of the boundary condition of zero wavefunction on the wall ensures the stability of the solution
near a solid wall in spite of the fact that the "domain wall" itself is
essentially unstable. In other words, the "hard wall" condition means that the wavefunction of BEC vanishes at the place where "a steep repulsion with a turn-on length smaller than the healing lengths and penetration depths of the condensates" exists \cite{Indekeu}. For example, potentials steeper than harmonic were prepared by using Laguerre-Gauss doughnut-shaped laser beams for a BEC container \cite{Doughnut}. An inhomogeneous stationary solution of GPE (the "domain wall") which coincides with the half of the dark soliton at rest (the kink soliton $\psi_{0}=\tanh(x)$, the distance $x$ is measured in the units of the healing length, see below) \cite{Pitaevskii1} may have one of its physical realization as a model for BEC  near a solid wall \cite{Kuznetsov} where localized Bogoliubov excitations were proposed to exist \cite{Surface of helium}.  However, an exact analytical solution for correspondent surface-bound excitations (if any) has not been found.

Although the stationary dark soliton in BEC was proven to be unstable with the "snake instability" \cite{Kuznetsov}, \cite{Muryshev}, the hard wall boundary condition may also approximate the sharpness of a self-bound potential at a free surface of liquid helium which was proven to be composed of nearly 100\% BEC that satisfies GPE \cite{Griffin}. One may consider the free kink wall of $\psi_{0}=\tanh(x)$ as a model for a free surface demanding only the topological stability of such a solution in which its nodal surface undergoes weak flexural oscillations. For this case the position of the "hard wall" is flexible (like an impenetrable membrane on the surface of helium II) and it imitates the free surface of the liquid. Then the role of a hard wall container is played by the liquid surface of helium II.

Here we consider the problem of the localized gapless excitation modes by finding analytical solutions of a matrix Schr\"{o}dinger equation being a slight (but important for what follows) modification of BdGE  \cite{A direct perturbation theory for dark solitons} as such as given in \cite{Kuznetsov}, \cite{Muryshev},  and \cite{Surface of helium}. The binding energy of localized excitations is of our concern in the present work.
Surprisingly, we found that the spectrum of surface excitations can be calculated analytically for any $k$ to be compared with the numerical results and with the analytical results obtained for limiting cases $k \rightarrow 0$ and $k \rightarrow \infty$. The natural limit of $k \rightarrow \infty$ which in the bulk BEC results in the energy spectrum $\varepsilon = {(\hbar k)^2}/ {2m}+\mu$ where $m$ is the mass of the boson and $\mu=gn_0$ is the chemical potential while $g$ and $n_0$ are the intensity of the repulsion and the BEC particle density, correspondingly, leads to $\varepsilon = {(\hbar k)^2}/ {2m}+\mu - \Delta$. The analytical approach  developed below for solving the Bogoliubov-de Gennes equations may turn useful for other applications.

\section{Basic equations}
GPE can be written as \cite{Ginzburg}:
\be
i\hbar\frac{\partial \psi}{\partial t}= - \frac{\hbar^2}{2m}\nabla^2 \psi + gn_0(|\psi|^2-1)\psi.
\label{eq:s0}
\ee
We introduce dimensionless quantities by measuring distances in the units of the healing length $\xi=\hbar/mc$ and the energy in the units of $gn_0=mc^2$ where $c=\sqrt{gn_0/m}$ is the sound velocity.
The stationary equation (\ref{eq:s0}) for the kink with the node at the position $x=0$ gives $\psi_0=\tanh(x)$ of the soliton .
We will disturb this solution to investigate its Bogoliubov excitations by presenting $\psi$ of Eq. (\ref{eq:s0}) as a sum of plane waves  \cite{Pitaevskii}:
 $
 \psi = \psi_0(x) + \vartheta( \vec{r}, t )$ with $\vartheta( \vec{r}, t ) = a_{\omega,\vec k}(x) \exp (i \vec{k} \vec{\varrho} -i \omega t) + b_{\omega,\vec k}^{*}(x) \exp (-i \vec{k} \vec{\varrho} + i \omega t)$,
where $\vec{r}=(x,\vec{\varrho})$, $\vec{\varrho}$ lies in the plane orthogonal to $x$ direction (we consider $x\geq 0$ and all the functions decaying exponentially in this direction), $\vec{k}$ is the wave vector along this plane and * denotes complex conjugation. We will suppress the indices and simplify the notation by using $a$ and $b$ instead of $a_{\omega,\vec k}(x)$ and $b_{\omega,\vec k}(x)$. Introducing the
functions $\psi_1 = a + b$ and $\psi_2 = a - b$, after linearizing  Eq. (\ref{eq:s0})  we get the
pair of coupled Schr\"{o}dinger equations  \cite{Surface of helium}:
\bea
-\frac{1}{2}\frac{d^{2}}{dx^{2}}\psi_{1}+(3\psi_{0}^{2}-1+\kappa^{2})\psi_{1} &=& \varepsilon \psi_{2}
\label{eq:s1} \\
-\frac{1}{2}\frac{d^{2}}{dx^{2}}\psi_{2}+(\psi_{0}^{2}-1+\kappa^{2})\psi_{2} &=& \varepsilon \psi_{1}
\label{eq:s2}
\eea
where $\kappa=\xi |\vec {k}|/\sqrt{2}$ and $\varepsilon=\hbar \omega/(mc^2)$. This pair of equations is identical to the corresponding Bogoliubov-de Gennes equations (see \cite{Quantum dark soliton} and \cite{A direct perturbation theory for dark solitons}) if one rewrites them for the functions $a$ and $b$, and also sets $\kappa=0$ and $\varepsilon=0$. As far as we know, Eqs. (\ref{eq:s1}),(\ref{eq:s2}) have never been solved before for arbitrary non-zero $\kappa$ and $\varepsilon$.

We find a formal general solution for these equations and illustrate its viability by obtaining the rigorous solution of the spectrum of localized phonons. The spectrum of bulk excitations can be easily found from (\ref{eq:s1}) and (\ref{eq:s2}) when neglecting the derivative terms far from the boundary $x=0$ to obtain the well-known Bogoliubov spectrum
$
\varepsilon_B = \kappa \sqrt{2+\kappa^2}
$
in the dimensionless form. For $\kappa \to 0$ it gives the bulk phonon $\varepsilon_B \approx \sqrt{2}\kappa+\kappa^3/(2\sqrt{2})$ and for $\kappa \to \infty$ $\varepsilon_B \approx \kappa^2 +1$ which is a free boson plus chemical potential. Any localized excitations should have the energy spectrum lying lower than the bulk one.

\section{Supersymmetry of BdGE}
It is interesting to note that Eqs. (\ref{eq:s1}), (\ref{eq:s2}) with $\varepsilon=0$ and  $\kappa= 0$ are parts of a supersymmetric Hamiltonian with zero ground state energy. Indeed, on introducing the matrix operator
\be
\hat{A}= \left( \begin{array}{cc}
-\frac{1}{\sqrt{2}}\frac{d}{dx} - \sqrt{2}\psi_{0} &  0\\
0 & -\frac{1}{\sqrt{2}}\frac{d}{dx}+ \frac{1}{\sqrt{2}}\frac{1-\psi_{0}^{2}}{\psi_{0}}  \end{array} \right) \label{eq:s2a}
\ee
so that the l.h.s. of Eqs. (\ref{eq:s1}), (\ref{eq:s2}) takes the form of a matrix Hamiltonian
\be
\hat H_{-} =\hat{A}^{\dag} \hat{A}= \left( \begin{array}{cc}
-\frac{1}{2}\frac{d^{2}}{dx^{2}}+3\psi_{0}^{2}-1&  0\\
0 & -\frac{1}{2}\frac{d^{2}}{dx^{2}}+\psi_{0}^{2}-1  \end{array} \right)
\label{eq:s2b}
\ee
with its partner Hamiltonian
\be
\hat H_{+} =\hat{A} \hat{A}^{\dag}= \left( \begin{array}{cc}
-\frac{1}{2}\frac{d^{2}}{dx^{2}}+\psi_{0}^{2}+1&  0\\
0 & -\frac{1}{2}\frac{d^{2}}{dx^{2}}+\frac{1-\psi_{0}^{2}}{\psi_{0}^{2}}  \end{array} \right)
\label{eq:s2c}
\ee
to produce a supersymmetric (SUSY) Hamiltonian
\be
\hat H_{SUSY} =\left( \begin{array}{cc}
\hat H_{-}&  0\\
0 & \hat H_{+}  \end{array} \right)
\label{eq:s2d}
\ee
that may canonically be expressed through the supercharges
\be
\hat Q = \left( \begin{array}{cc}
0 &  0\\
\hat{A} & 0 \end{array} \right) ,
\hat Q^{\dag} =\left( \begin{array}{cc}
0 &  \hat{A}^\dag\\
0 & 0 \end{array} \right)
\label{eq:s2e}
\ee
as anticommutator 
\be
\hat H_{SUSY}=\{Q,Q^\dag \};
\hat Q^2=0, (\hat Q^{\dag})^2=0 .
\label{eq:s2f}
\ee
The supersymmetry is explicitly broken at $\varepsilon>0$ and  $\kappa > 0$ which eventually leads to splitting the degenerate zero ground state into two gapless excitations (a "light" one with $\varepsilon \propto \kappa $ and a "heavy" one with $\varepsilon \propto \kappa^{3/2} $)\cite{Surface of helium}, both bound to the wall.

\section{Boundary conditions}
The boundary conditions for $\psi_1$ and $\psi_2$ can be of two kinds \cite{Surface of helium}. At the node of the kink $\psi=0$ that is both ${\rm Im}~ \psi=0$ and ${\rm Re}~ \psi=0$ therefore  $\psi_2=0$ and $\psi_1=0$. However, for $\psi_1$ the additional possibility exists. Indeed, for $\kappa=0$ and $\varepsilon=0$ Eqs. (\ref{eq:s1}),(\ref{eq:s2}) have the solutions $\psi_1^0=1-\psi_0^2$ and $\psi_2^0=\psi_0$ the first of which is the so-called "zero mode"  \cite{Surface of helium}, \cite{Quantum dark soliton} or Goldstone gapless mode corresponding to a translation of the kink $\psi_0$ as a whole along $x$, being the derivative of the kink $\psi_0$: $\psi_0(x+\delta x)\approx \psi_0(x)+\psi_1^0 \delta x$. Thus the condition ${\rm Re}~ \psi=0$ turns into $\psi_0' \delta x(\vec \varrho, t)+ {\rm Re}~ \vartheta (\vec r, t)=0$ which determines the shape of the loci of nodes $\delta x(\vec \varrho, t)$ (the shape of the surface). The derivative of such a mode with respect to $x$ is zero at $x=0$. Thus the mode with the mixed boundary conditions
$\frac{d}{dx}\psi_1\mid_{x=0}=0 $ and $\psi_2\mid_{x=0}=0$
would allow the rippling of the soliton and will be called the "ripplon" mode (predicted in \cite{Surface of helium}). As we shall see below its energy spectrum at low $\kappa $ coincides with the one for the capillary wave. The mode with the zero boundary conditions $\psi_1\mid_{x=0}=0$ and $\psi_2\mid_{x=0}=0$
which correspond to a flat boundary of the solid wall will be called the "surface phonon" mode (as far as its spectrum starts linear) and was predicted in \cite{Surface of helium}. Finally, the very condition of the solid wall excludes the possible solution $x\psi_0 -1$ \cite{Muryshev}  of Eq. (\ref{eq:s2}) at $\kappa=0$, $\varepsilon=0$ which could be responsible for the "snake" instability of the "domain wall" and which does not satisfy the zero boundary conditions.

\section{Long-wavelength approximation solution}
\begin{figure}[h]
  \includegraphics[width=1\linewidth,clip=]{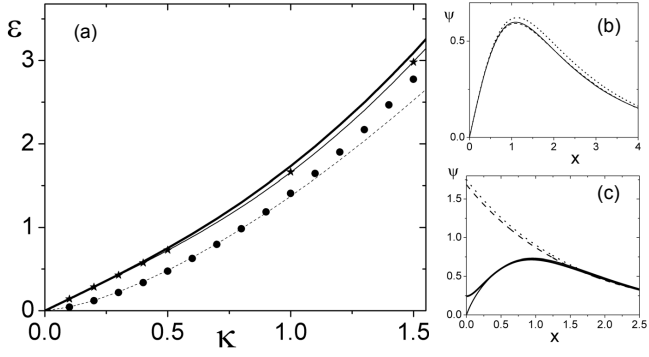}\\
  \caption{The dimensionless spectra of the elementary excitations vs the dimensionless wavenumber and wavefunctions. (a) The thick solid line is the Bogoliubov bulk excitation spectrum, the dashed line is the capillary wave spectrum (\ref{eq:rip}), the circles mark the  spectrum of the ripplon calculated by numerical solving Eqs. (\ref{eq:s1}),(\ref{eq:s2}), the stars mark the numerical spectrum of the surface phonon and the thin line is the exact solution (\ref{eq:s18}). As it should be its energy lies higher than the energy of the ripplon because of two nodes in the zero boundary conditions; (b) the numerical wavefunctions of the surface phonon at $\kappa=3.5$ are shown with dashed ($\psi_1$) and dot ($\psi_2$) lines together with $\psi_\infty$ (solid line) which they approach; (c) the numerical wavefunctions of the ripplon mode at $\kappa=5$. It is seen that $\psi_1$ (thick solid line) lies very close to $\psi_2$ except the coordinate origin where $\psi_1$ has zero derivative. The dashed and dot lines show the asymptotic  $\exp(-\alpha_2 x)$ for $\psi_{1,2}$.}\label{fig: rpl}
\end{figure}

First consider the case of $\kappa \to 0$.  In case of the ripplon spectrum we find $\psi_{1,2}$ as a series in $\varepsilon$:
$\psi_1\approx \psi_1^0 + \varepsilon \psi_1^1 + \mathcal{O}(\varepsilon^2)$ and
$\psi_2\approx \psi_2^0 + \varepsilon \psi_2^1 + \mathcal{O}(\varepsilon^2)$.
A zero approximation is the solution of homogeneous equations Eqs. (\ref{eq:s1}), (\ref{eq:s2}) with $\varepsilon=0$. The solutions can be found for any $\kappa$ (this can be verified by the direct substitution):
\bea
\label{eq:z1}
\psi_1^0=A \exp(-\alpha_1x)\left(\frac{\alpha_1^2-1}{3}+\alpha_1\psi_0 + \psi_0^2\right)\\
\label{eq:z2}
\psi_2^0=B \exp(-\alpha_2x)(\psi_0 + \alpha_2),
\eea
where $\alpha_1=\sqrt{2}\sqrt{2+\kappa^2}$ and $\alpha_2=\sqrt{2}\kappa$.
To find $\psi_1^1$ and $\psi_2^1$ we have to solve the inhomogeneous equations that follow from Eqs. (\ref{eq:s1}), (\ref{eq:s2}) where $\kappa=0$:
\bea
\label{eq:ss1}
-\frac{1}{2}\frac{d^{2}}{dx^{2}}\psi_{1}^1+(3\psi_{0}^{2}-1)\psi_{1}^1 &=& B\psi_{0} \\
 \label{eq:ss2}
-\frac{1}{2}\frac{d^{2}}{dx^{2}}\psi_{2}^1+(\psi_{0}^{2}-1)\psi_{2}^1 &=&  A(1 - \psi_{0}^2).
\eea
With the help of the Green functions of the homogeneous equations the inhomogeneous solutions are found as:
\bea
\label{eq:az1}
\psi_1^1=B\frac{1}{2}(\psi_0+x(1 - \psi_0^2))\\
\label{eq:az2}
\psi_2^1=-A.
\eea
Finally, the derivative with respect to $x$ of $\psi_1$ at $x=0$ is found from Eqs. (\ref{eq:z1}), (\ref{eq:az1}) to be $\psi_1'=A\alpha_1(2-\alpha_1)(2+\alpha_1)/3 + \varepsilon B$, which according to the mixed boundary conditions should be zero together with $\psi_2=-A\varepsilon + B \alpha_2$, as it follows from Eqs. (\ref{eq:z2}), (\ref{eq:az2}). The zero determinant with respect to $A$ and $B$
\be
\det \left( \begin{array}{cc}
\alpha_1(2-\alpha_1)(2+\alpha_1)/3 & \varepsilon\\
-\varepsilon & \alpha_2 \end{array} \right)=0
\label{eq:zz3}
\ee
gives the ripplon spectrum taking into account that $\alpha_1 \approx 2 + \kappa^2/2$ for $\kappa\to 0$ and retaining only the lowest power of $\kappa$
\be
\varepsilon=\sqrt{\frac{4\sqrt{2}}{3}}\kappa^{3/2}.
\label{eq:rip}
\ee
Spectrum (\ref{eq:rip}) is shown in Fig. \ref{fig: rpl}. Note that the localization of the ripplon at low $\kappa$ is governed by $\alpha_2=\sqrt{2}\kappa$ which is compared below with the numerical solution for $\alpha_2$. As it was already noted in \cite{Surface of helium} the spectrum (\ref{eq:rip}) exactly coincides with the well-known expression for the frequency of the capillary waves multiplied by the Planck constant when written in a dimensional form: $\varepsilon = \hbar \sqrt{\sigma/mn_0}~ k^{3/2}$ where $\sigma=2/3~\hbar c n_0$ is the surface energy of the stationary soliton $\psi_0$ \cite{Ginzburg}. In fact, $\sigma$ is exactly the half of the energy of the dark soliton at rest (see Eq. (5.59) in \cite{Pitaevskii1}).

The zero boundary conditions lead to surface phonons for $\kappa\to 0$ \cite{Surface of helium} and below we obtain the whole spectrum analytically. Here we only mention  that $\alpha_2$ for phonons at low $\kappa$ is proportional to $\kappa^2$ which indicates much weaker localization, in contrast to the ripplons.

\section{Short-wavelength approximation solution}
For the case $k \rightarrow \infty$ introduce function $\chi$ and constant $\Delta$ so that $\psi_{1}=\psi_{2}+\chi/k^2$, $\psi_{2}=\psi_\infty$ and $\epsilon=\kappa\sqrt{\kappa^2+2}-\Delta \asymp \kappa^2+1-\Delta $. Then Eqs. (\ref{eq:s1}), (\ref{eq:s2}) turn into
 \bea
 \label{eq:s7}
-\frac{1}{2}\frac{d^{2}}{dx^{2}}\psi_{\infty}+(3\psi_{0}^{2}-2+\Delta)\psi_{\infty} &=& - \chi\\
\label{eq:s8}
-\frac{1}{2}\frac{d^{2}}{dx^{2}}\psi_{\infty}+(\psi_{0}^{2}-2+\Delta)\psi_{\infty} &=& \chi
\eea
which after adding and subtracting both equations lead to
\be
\psi_\infty = (1-\psi_0^2)^{\alpha_{\infty} /2} {}_{2}F_{1}(\alpha_{\infty} - s,\alpha_{\infty}+s+1, \alpha_{\infty} + 1, \frac{1-\psi_0}{2} )
\label{eq:s10}
\ee
where the hypergeometric function contains $\alpha_{\infty} = \sqrt{2\Delta}$ and $s=(\sqrt{17}-1)/2$ is the solution of the equation $s(s+1)=4$ (see \cite{Landau}) and $\chi=-\psi_0^2 \psi_\infty$. The boundary condition $\psi_2=0$ at $x=0$ gives the equation
\begin{multline}
{}_{2}F_{1}(\alpha_{\infty} - s,\alpha_{\infty}+s+1, \alpha_{\infty} + 1, \frac{1}{2} ) = \\ \frac{\Gamma(\frac{1}{2})\Gamma(\alpha_{\infty} +1)}{\Gamma(\frac{1}{2}(1+\alpha_{\infty}-s))\Gamma(\frac{1}{2}(2+\alpha_{\infty}+s))}= 0.
\label{eq:s11}
\end{multline}
which fulfils for $1+\alpha_{\infty} -s=0$ and therefore $\alpha_{\infty}=(\sqrt{17}-3)/{2}\approx 0.562 $ while $\Delta=\alpha_\infty^2/2\approx 0.158 $. Finally, the hypergeometric function in (\ref{eq:s10}) reduces to $\psi_0$ so that $\psi_\infty=\psi_0(1-\psi_0^2)^{\alpha_{\infty}/2}=\tanh(x)/\cosh(x)^{\alpha_{\infty}}$ (see Fig.\ref{fig: rpl} (b)) and $\chi=-\psi_0^3(1-\psi_0^2)^{\alpha_{\infty}/2}$, therefore, both $\psi_2$ and $\psi_1$ satisfy the zero boundary conditions. Analogously, one can show (we skip this calculation which makes use of the known solution of homogeneous equations (\ref{eq:z1}), (\ref{eq:z2}) to satisfy the boundary conditions) that the function $\psi_\infty$ is also the limiting function for large $\kappa$ for the mixed boundary conditions so that the difference between the functions is seen only in the close proximity to the boundary at the distance $1/\kappa$ (see Fig.\ref{fig: rpl}(c) where near $x=0$  $\psi_1$ deviates from $\psi_2$ and meets the $\psi$ axis with zero derivative).
Thus  the binding energy of the  excitation localized near the surface behaves universally as well as its wavefunction does. In dimensional units the binding energy is $0.158 mc^2 \approx 4$ K .

\section{Exact solution of BdGE}
Let us now find the exact solution of Eqs. (\ref{eq:s1}), (\ref{eq:s2}) at arbitrary $\kappa$. To do this let us transform these equations into a single matrix hypergeometric equation. Matrix generalizations of both hypergeometric function and gamma function were shown to be mathematically correct (see Refs. (\cite{Matrix valued}) and (\cite{Gamma and Beta})). Introducing $z=(1-\psi_0)/2$ and $\psi_{1,2}=(1-\psi_0^2)^{\alpha/2} \phi_{1,2}$ we rewrite Eqs. (\ref{eq:s1}), (\ref{eq:s2})~into
\begin{multline}
z(1-z)\frac{d^2\phi_1}{dz^2} + (\alpha+1 - 2(\alpha+1)z)\frac{d\phi_1}{dz} +( 6 - \alpha(\alpha + 1))\phi_1 \\
+\frac{1}{2z(1-z)}(\frac{\alpha^2}{2} - 2- {\kappa^2})\phi_1={\varepsilon}\frac{\phi_2}{2z(1-z)}\\
z(1-z)\frac{d^2\phi_2}{dz^2} + (\alpha + 1 - 2(\alpha+1)z)\frac{d\phi_2}{dz} + (2 - \alpha(\alpha + 1))\phi_2  \\
+\frac{1}{2z(1-z)}(\frac{\alpha^2}{2} - {\kappa^2})\phi_2={\varepsilon}\frac{\phi_1}{2z(1-z)}.
\label{eq:s12}
\end{multline}

To turn (\ref{eq:s12}) into a matrix hypergeometric equation we introduce the vector-function $\hat{\Phi}$, the unit matrix $\hat {1}$ and the matrices
\bea
\hat {\Phi} =  {\left( \begin{array}{c}
\phi_1 \\
\phi_2 \end{array} \right)} \\
\hat {\alpha}^2= 2\left( \begin{array}{cc}
2+{\kappa^2} &  {\varepsilon}\\
{\varepsilon} & {\kappa^2} \end{array} \right)\\
\hat {c}=\hat{\alpha}+\hat{1}\\
 \label{eq:sa13}
 \hat{1}+\hat{a}+\hat{b}=2(\hat{\alpha}+\hat{1})\\
 -\hat{a}\hat{b}= \left( \begin{array}{cc}
6 &  0\\
0 & 2 \end{array} \right)- \hat{\alpha}^2-\hat{\alpha}.
\label{eq:s13}
\eea
The matrix $\hat{\alpha}$ can be obtained as a square root of $\hat{\alpha}^2$ which gives
\be
\hat{\alpha}= \sqrt{\hat {\alpha}^2}= {2}\left( \begin{array}{cc}
r &  l\\
l & p \end{array} \right)
\label{eq:s13b}
\ee
where $r=\sqrt{1+\kappa^2/2-l^2}$, $ p= \sqrt{\kappa^2/2-l^2}$, and $ l={\varepsilon}\sqrt{\kappa^2+1-(\kappa^2(\kappa^2+2)-\varepsilon^2)^{1/2}}/(2\sqrt{\varepsilon^2+1})
$. The positive eigenvalues of the matrix $\hat \alpha$ are
\be
 \alpha_{1,2} = \sqrt{2}(1+k^2 \pm \sqrt{\varepsilon^2+1})^{1/2}
\label{eq:sa14}
\ee
and $\exp (-\alpha_2 x)$ determines the asymptotic decay  of $\psi_1$ and $\psi_2$ as $x\to \infty$.

After introducing the matrices, Eq. (\ref{eq:s12}) becomes
\be
z(1-z)\hat{\Phi}''+(\hat{c}-(\hat{1}+\hat{a}+\hat{b})z)\hat{\Phi}' -\hat{a}\hat{b}\hat{\Phi}=0,
\label{eq:s14}
\ee
where primes mean differentiation with respect to $z$.

\begin{figure}[h]
\centerline{\includegraphics[width=1\linewidth,clip=]{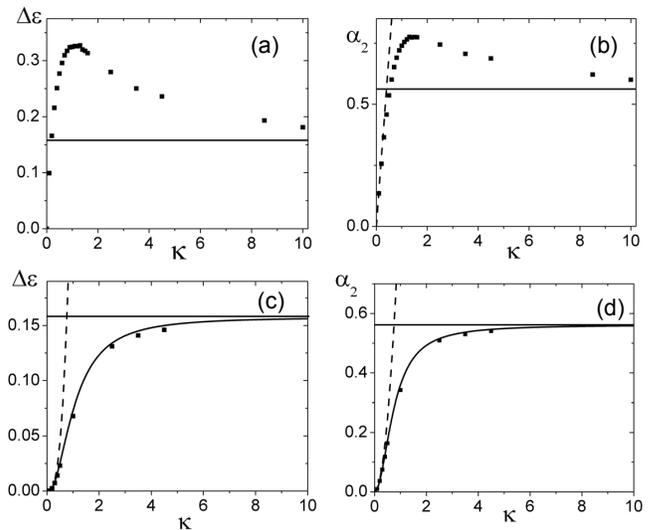}}
\caption{(a)~ The dimensionless binding energy of the ripplon vs the dimensionless wavenumber. The squares show the results of the numerical solution of Eqs. (\ref{eq:s1}), (\ref{eq:s2}) with the mixed boundary conditions. The horizontal line is drawn at the energy $\Delta = 0.158$;~(b) The squares show the decay parameter $\alpha_2$ for the ripplon obtained with  Eq.(\ref{eq:sa14}) . The horizontal line is drawn at $\alpha_\infty = 0.568$. The dashed line is the analytical solution $\alpha \approx \sqrt{2}\kappa$ at low $\kappa$;~(c)~The  binding energy of the surface phonon vs the wavenumber. The solid line is the result of the exact solution given by Eq. (\ref{eq:s18}). The squares show the results of the numerical solution of Eqs. (\ref{eq:s1}),(\ref{eq:s2}) with the zero boundary conditions. The horizontal line is drawn at the energy $\Delta$. The dashed line is  $\kappa^3/(2\sqrt{2})$ that follows from the low $\kappa$ behavior (see text); (d)~ The solid line is  the result of the exact solution for $\alpha_2$ given by Eqs. (\ref{eq:s18}),(\ref{eq:sa14}). The squares show the results of the numerical solution of Eqs. (\ref{eq:s1}), (\ref{eq:s2}). The horizontal line is drawn at $\alpha_\infty$. The dashed line is $\kappa^2$ predicted in \cite{Surface of helium} (see text).}
\label{fig:phonenerg}
\end{figure}

Equation (\ref{eq:s14}) has a formal solution as the matrix-valued hypergeometric function \cite{Matrix valued}
\be
{}_2F_1(a,b,c,z)=\sum_{n\geq 0} \frac{z^n}{n!}(\hat{a},\hat{b},\hat{c})_n ,
\label{eq:saa14}
\ee
where $(\hat{a},\hat{b},\hat{c})_{0}=1; $
\begin{multline}
(\hat{a},\hat{b},\hat{c})_{m+1} =
(\hat{c}+m)^{-1}(\hat{a}+m)(\hat{b}+m)\times \\
(\hat{c}+m-1)^{-1}(\hat{a}+m-1)(\hat{b}+m-1)\ldots \hat{c}^{-1}\hat{a}\hat{b}.
\label{eq:s15}
\end{multline}
We use this formal solution to obtain the spectrum of the phonon localized near the soliton. The boundary condition at $x=0$ (that is at $z=1/2$) will be fulfilled when $\hat \Phi=0$. As far as the hypergeometric function at $z=1/2$ can be expressed through the matrix gamma function (\cite{Gamma and Beta})  (because $\hat {c}=(\hat{1}+\hat{a}+\hat{b})/2$, see Eq.(\ref{eq:sa13})) so that
\be
{}_{2}F_{1}(\hat{a},\hat{b}, \hat{c}, \frac{1}{2} ) = \frac{\Gamma(\frac{1}{2})\Gamma(\hat{c})}{\Gamma\left(\frac{1}{2}(1+\hat{a})\right) \Gamma\left(\frac{1}{2}(1+\hat{b})\right)}
\label{eq:s16}
\ee
then the condition of zero $\Phi$  means that the matrix (\ref{eq:s16}) has a zero eigenvalue, that is the determinant of (\ref{eq:s16}) should be zero. In turn, matrix gamma functions could be presented as the products of matrices \cite{Gamma and Beta}
\be
\Gamma(\hat M)=\lim_{n\to \infty} (n-1)! n^{\hat M}[\hat M (\hat M+\hat 1)\dots (\hat M+n \hat 1) ]^{-1} ,
\ee
the most relevant for the zero determinant should be the determinant of either $\hat 1 +\hat a$ or $\hat 1 +\hat b$. In fact one can prove that the determinant of each of them gives the same spectrum. However an analytical solution for the matrix equations (\ref{eq:sa13}), (\ref{eq:s13}) for $\hat a$ and $\hat b$ is difficult. To obtain the analytical results we will instead utilize the product $\hat{P}=((\hat{1}+\hat{a})(\hat{1}+\hat{b}))/2=(\hat{1}+\hat{a}+\hat{b}+\hat{a}\hat{b})/2$ that uses already calculated matrices. It is easy to obtain $\hat{P}$ from (\ref{eq:sa13}),(\ref{eq:s13}) and (\ref{eq:s13b})
\be
\hat{P}= \left( \begin{array}{cc}
3r+\kappa^2 &  3l +\varepsilon\\
3l+\varepsilon & 3p + \kappa^2 \end{array} \right).
\label{eq:s17}
\ee
Finally, the spectrum of the surface phonons is determined by the equation
\be
\det \hat{P} =(3r+\kappa^2)(3p + \kappa^2)-(3l +\varepsilon)^2 =0.
\label{eq:s18}
\ee
One can make sure that Eq. (\ref{eq:s18}) reproduces the spectrum calculated before at $\kappa \to 0$ and $\kappa \to \infty$. Indeed, at $\kappa \to 0$ the spectrum is $\varepsilon = \sqrt{2}\kappa + \mathcal{O}(\kappa^5)$ so that the $\kappa^3$ term is missing as it was predicted in \cite{Surface of helium}, while the bulk phonon starts with higher energy as $\varepsilon_{B} = \sqrt{2}\kappa + \kappa^3/(2\sqrt{2}) + \mathcal{O}(\kappa^5)$. Let us define the absolute value of the binding energy as $\Delta \varepsilon = \varepsilon_{B}-\varepsilon$. Then it starts as $\kappa^3/(2\sqrt{2})$ (see Fig. \ref{fig:phonenerg}(c)). Now let us consider the other limit $\kappa \to \infty$. It is easy to see that seeking the solution in the form $\varepsilon \asymp \kappa^2+1-\Delta$  leads to $r=p \asymp \kappa/2 + \sqrt{2\Delta}/4$ and $l\asymp \kappa/2 - \sqrt{2\Delta}/4$ which after substitution into Eq. (\ref{eq:s18}) give
$ 2\Delta+3\sqrt{2\Delta}-2=0. $
It has the same root as we found before from Eq. (\ref{eq:s11}) $\sqrt{2\Delta}=(\sqrt{17}-3)/2=\alpha_{\infty}\approx 0.562$ and therefore $\Delta \varepsilon_{\infty}=\Delta=\alpha_{\infty}^2/2\approx 0.158$. Such a coincidence with exact asymptotic results found before brings confidence to Eq. (\ref{eq:s18}) which is enhanced by a good correspondence with the results obtained by direct numerical solution of the coupled Shr\"{o}dinger equations (Fig. \ref{fig:phonenerg}). Note that the slower decay exponent $\alpha_2$ can be approximated by a simple expression $\alpha_2=\kappa^2/(1+\kappa^2/\alpha_\infty)$ that fits the exact expression of Eq. (\ref{eq:sa14}) with $\varepsilon$ from the exact solution of Eq. (\ref{eq:s18}) within $0.2 \%$. We plot the decay exponential in Fig. \ref{fig:phonenerg}(d) together with the results of the numerical solution. Unfortunately, the case of mixed boundary conditions for ripplons can not be treated in the same way.
\section{Conclusion}
In conclusion, we have found the analytic spectra and wavefunctions of localized Bogoliubov elementary excitations existing near the inhomogeneous stationary solution of the Gross-Pitaevskii equation which may represent a physical  model for the surface of liquid helium. In this status our solutions predict surface modes - ripplons and surface phonons - that may contribute to the thermodynamics of the surface at low temperatures \cite{Surface of helium}. We believe that our results and the method for exact solving the Bogoliubov-de Gennes equations could be useful in more sophisticated cases involving solitons.

\section{Acknowledgement}
This work was supported by the Global Frontier Center for Multiscale Energy Systems funded by National Research Foundation under the Ministry of Education, Science and
Technology (2011-0031561). Financial support from BK21 program and WCU (World Class University) multiscale mechanical design program (R31-2008-000-10083-0) through
the Korea Research Foundation is gratefully acknowledged.

\end{document}